\documentclass[pre,aps ,superscriptaddress,showpacs]{revtex4-1}
\usepackage{natbib}
\usepackage{amsmath}
\usepackage{epsfig}
 \usepackage{color}

\begin{document}

\title{Quantum corrections of work statistics in closed quantum systems}
\author{Zhaoyu Fei}
\author{H. T. Quan}
\email[Email address: ]{htquan@pku.edu.cn }
\affiliation{School of Physics, Peking University, Beijing 100871, China }
\author{Fei Liu}
\email[Email address: ]{feiliu@buaa.edu.cn}
\affiliation{School of Physics and Nuclear Energy Engineering, Beihang University, Beijing 100191, China}

\date{\today}

\begin{abstract}
{We investigate quantum corrections to the classical work characteristic function (CF) as a semiclassical approximation to the full quantum work CF. In addition to explicitly establishing the quantum-classical correspondence of the Feynman-Kac formula, we find that these quantum corrections must be in even powers of $\hbar$. Exact formulas of the lowest corrections ($\hbar^2$) are proposed, and their physical origins are clarified. We calculate the work CFs for a forced harmonic oscillator and a forced quartic oscillator respectively to illustrate our results. }

\end{abstract}
\maketitle

\section{Introduction}
\label{section1}
Recently, the statistics of quantum work has attracted considerable attention~\cite{Talkner2007,Campisi2011,Esposito2009,Liu2018}. This issue was initially motivated by theoretical efforts on extending classical fluctuation theorems~\cite{Bochkov1977,Evans1993,Gallavotti1995,Jarzynski1997,Kurchan1998,
Lebowitz1999,Maes1999,Crooks2000,Hatano2001,Seifert2005,Sagawa2010} into quantum regimes. The practical feasibility of manipulating and/or controlling the energy of small quantum systems further boosted these research interests~\cite{ShuomingAn2015,Rossnagel2016,Batalhao2014,Xiong2018}.

In closed quantum systems, quantum work is defined by the two energy measurement scheme (TEM)~\cite{Kurchan2000,Tasaki2000}. Although this definition has been criticized because of the destruction of possible initial coherence~\cite{Perarnau-Llobet2017,Allahverdyan2005}, several works support its justification from the aspect of the quantum-classical correspondence principle~\cite{Jarzynski2015,Zhu2016,Garcia-Mata2017}. Obviously, this classical correspondence is not the ultimate goal of these studies; the quantum characteristics of quantum work are the main concerns. A possible improvement to full classical work statistics is to develop semiclassical approaches. In addition to academic interest, we expect these semiclassical approaches to provide practical methods to compute complex quantum work statistics of general quantum systems. In this paper, we present such an approach. We follow the idea of Wigner~\cite{Wigner1932} and represent an evolution equation for the characteristic function (CF) of quantum work in the phase space of a system. By expanding the equation in powers of the Planck constant, $\hbar$, the classical work statistics and their quantum corrections are clearly revealed.

This paper is organized as follows. In Sec.~(\ref{section2}), we briefly review the CF method of quantum work in closed quantum systems. In Sec.~(\ref{section3}), a quantum-classical correspondence of the Feynman-Kac formula is established. In Sec.~(\ref{section4}), we present the lowest quantum corrections to the classical CF. Forced harmonic and quartic oscillators are used to illustrate our results in Sec.~(\ref{section5}). Sec.~(\ref{section6}) presents the conclusion.

\section{Overview of the CF method}
\label{section2}
Let us begin with the quantum work definition of a closed quantum systems with Hamiltonian $\hat H(t)$. Throughout this paper, all operators are denoted by hat in order to distinct them from their classical correspondences in the phase space of a system. Because the aim of this work is to illustrate the idea, our discussion is limited to the simplest single particle and one-dimensional situations. According to the two-energy-measurement (TEM) scheme~\cite{Kurchan2000,Tasaki2000}, given that the instantaneous energy eigenvectors and eigenvalues of the Hamiltonian are
\begin{eqnarray}
\label{eigenvectorandvaluesofclosedHamiltonian}
\hat H(t)|\varepsilon_n(t)\rangle=\varepsilon_n(t) |\varepsilon_n(t)\rangle,
\end{eqnarray}
the quantum work done by various external agents on the quantum system is defined as $W_{nm}=\varepsilon_n(t)-\varepsilon_m(0)$.
By repeating this measurement scheme many times, one can construct the probability distribution of the stochastic work as
\begin{eqnarray}
\label{distributioninclusiveworkclosedsystem}
P(W)=\sum_{n,m} \delta( W-W_{nm}) \left |\langle \epsilon_n(t)|U(t)|\epsilon_m(0)\rangle \right |^2 P_m(0),
\end{eqnarray}
where $U(t)$ is the time evolution operator of the system, and $P_m(0)$ is the probability of finding the system of the eigenvector $|\epsilon_m(0)\rangle$ at time $0$. We assume that the system is initially in the thermal equilibrium state, that is, $P_m(0)=\exp[-\beta\epsilon_m(0)]/Z(0)$, where
$Z(0)$ is the partition function at time $0$ and is equal to ${\rm Tr}\{\exp[-\beta \hat H(0)]\}$, and $\beta$ is the inverse temperature. Because the Dirac function is involved, Eq.~(\ref{eigenvectorandvaluesofclosedHamiltonian}) is not the most convenient form to analyze the statistical properties of the work distribution. An alternative way is to resort to its Fourier transform or CF~\cite{Kurchan2000,Talkner2007,Esposito2009,Campisi2011,Liu2018}, and it can be re-expressed by taking the trace over an operator:
\begin{eqnarray}
\label{CFworkclosedsystem}
\Phi(\eta)&=&{\rm Tr}\left[ e^{i\eta\hat{H}(t)} U(t)e^{-i\eta \hat{H}(0)}\rho_0U^\dag(t)\right]\equiv{\rm Tr}[\hat{K}(t)].
\end{eqnarray}
We call $\hat K(t)$ the work characteristic operator (WCO)~\cite{Liu2018}. It is easy to prove that the operator satisfies the following evolution equation~\cite{Liu2012}:
\begin{eqnarray}
\label{timeevolutionequationclosedsystework}
\partial_t \hat{K}(t) =&&\frac{1}{i\hbar}[\hat{H}(t),\hat{K}(t)]+ \left[\partial_t e^{i\eta \hat{H}(t)}\right ] e^{-i\eta \hat{H}(t)} \hat{K}(t) \nonumber \\
\equiv&&\frac{1}{i\hbar} [\hat{H}(t),\hat{K}(t)]+ {\hat{\Omega}}(t)\hat{K}(t).
\end{eqnarray}
Note that the initial condition is $\hat{K}(0)=\exp[-\beta \hat H(0)]/{Z(0)}$. The collection of Eqs.~(\ref{distributioninclusiveworkclosedsystem})-(\ref{timeevolutionequationclosedsystework}) is called the quantum Feynman-Kac (FK) formula since it is fully consistent with the spirit of the original paper by Kac~\cite{Kac1949} on the establishment of the CF method for evaluating the distributions of the classical stochastic functional~\cite{Liu2018}.

\section{Quantum-classical correspondence of the FK formula}
\label{section3}
Following the idea of Wigner~\cite{Wigner1932}, we reformulate Eq.~(\ref{timeevolutionequationclosedsystework}) in the phase space representation~\cite{Wigner1932,HILLERY1984121,Polkovnikov2010}: let $K(z,t)$ be the Weyl symbol of $\hat K(t)$, where $z=(x,p)$ is the phase point, and $x$ and $p$ are the position and momentum of the particle, respectively. Then,
\begin{eqnarray}
\label{equationmotioninWeylsymbol}
{\partial_t K} =-\frac{2}{\hbar} H\sin\left(\frac{\hbar\Lambda}{2}\right)K+\Omega
\exp\left(\frac{-i\hbar}{2}\Lambda \right)K ,
\end{eqnarray}
where the symplectic operator is~\cite{HILLERY1984121}
\begin{eqnarray}
\Lambda=\overleftarrow{\partial_p}\overrightarrow{\partial_x}-\overleftarrow{\partial_x}\overrightarrow{\partial_p},
\end{eqnarray}
and the arrows indicate which direction the derivatives act upon. $\Omega(z,t)$ is the Weyl symbol of $\hat \Omega$:
\begin{eqnarray}
\label{OmegaoperatorWeylsymbol}
\Omega(z,t)=\left[ \partial_t e^{i\eta {\hat H}(t)} \right]_w\exp\left(\frac{-i\hbar}{2}\Lambda \right) \left[ e^{-i\eta {\hat H}(t)}\right]_w.
\end{eqnarray}
Here the subscript ``$w$" is used to indicate the Weyl symbols of these exponential operators. If one solves Eq.~(\ref{equationmotioninWeylsymbol}), the CF of the quantum work is then evaluated directly as
\begin{eqnarray}
\Phi(\eta)=\int_{-\infty}^{+\infty} dz K(z,t).
\end{eqnarray}

Eq.~(\ref{OmegaoperatorWeylsymbol}) appears to be very complicated. Importantly, Wigner~\cite{Wigner1932} obtained the Weyl symbol of the exponential Hamiltonian by expanding it in powers of $\hbar$ when investigating the quantum corrections to the classical thermodynamical quantities~\footnote{Different from this case, Wigner studied the expansion of $\exp(-\beta \hat H)$ in powers of $\hbar$. However, his formulas remain valid, and we simply replace $\beta$ by $i\eta$ in his results. }:
%%%Quality Control Editor - Please ensure that the intended meaning has been maintained in the above edit.
\begin{eqnarray}
\label{exponentialHamiltonianWeylsymbol}
\left[ e^{-i\eta \hat H(t)}\right]_w=e^{-i\eta H(z,t)}\left[ 1 + (i\hbar)^2 f(i\eta,z,t)+o(\hbar^2) \right],
\end{eqnarray}
where
\begin{eqnarray}
\label{ffunction}
f(i\eta,z,t)=\frac{(i\eta)^2 }{8m}\left[ {\partial_x^2}U - \frac{i\eta}{3}\left({\partial_x }U  \right)^2  - \frac{ i\eta}{3m} p^2  {\partial_x^2}U \right].
\end{eqnarray}
For simplicity, we restrict our discussion to a simple system The Hamiltonian of the quantum system is,
\begin{eqnarray}
\label{Hamiltonian}
\hat H(t)=\frac{\hat p^2}{2m}+ U(\hat x, t),
\end{eqnarray}
where $m$ is the mass of the single particle system. Substituting Eqs.~(\ref{OmegaoperatorWeylsymbol}) and (\ref{exponentialHamiltonianWeylsymbol}) into the right-hand side of Eq.~(\ref{equationmotioninWeylsymbol}) and expanding it to the second power of $\hbar$, we have
\begin{eqnarray}
\label{equationmotioninphasespacetillsecondorder1}
{\partial_t K} &=&-H \Lambda K + i\eta {\partial_t H}  K +\frac{i\hbar}{2}\left[ (i\eta)^2 \left( \partial_t H \Lambda H \right)-i\eta {\partial_t H} \Lambda\right] K +\left(i\hbar\right)^2\left(\cdots  K\cdots\right)+\cdots%{\cal O}(i\hbar)^3.
\end{eqnarray}
The exact expression of $(\cdots)$ is presented in Appendix A. We do not include terms with higher powers of $\hbar$, which can be calculated in a similar way in principle. To investigate the quantum-classical correspondence and possible quantum corrections, we expand $K$ in powers of $\hbar$ as follows:
\begin{eqnarray}
\label{e13}
K=K^{(0)}+(i\hbar) K^{(1)}+ (i\hbar)^2 K^{(2)}+\cdots.
\end{eqnarray}
Substituting Eq.~(\ref{e13}) into Eq.~(\ref{equationmotioninWeylsymbol}) and collecting all terms with the same powers of $\hbar$, we obtain
\begin{eqnarray}
\label{equationmotioninphasespace0order}
{\partial_t K^{(0)}} &=&-H \Lambda K^{(0)} + i\eta {\partial_t H} K^{(0)},\\
\label{equationmotioninphasespace1order}
{\partial_t K^{(1)}}&=&-H \Lambda K^{(1)} + i\eta {\partial_t H}  K^{(1)} + \frac{1}{2}\left[ (i\eta)^2 \left( {\partial_t H} \Lambda H \right) -i\eta {\partial_t H} \Lambda\right] K^{(0)},\\
\label{equationmotioninphasespace2order}
{\partial_t K^{(2)}}&=&-H \Lambda K^{(2)} + i\eta {\partial_t H}  K^{(2)} +\frac{1}{2}\left[ (i\eta)^2 \left({\partial_t H} \Lambda H \right) - i\eta {\partial_t H} \Lambda\right] K^{(1)}+\left(\cdots K^{(0)}\cdots\right) .
\end{eqnarray}
The initial conditions are
\begin{eqnarray}
\label{initialconditionK0}
K^{(0)}(z,0)&=&P_{eq}(\beta,z,0),\\
\label{initialconditionK1}
K^{(1)}(z,0)&=&0,\\
\label{initialconditionK2}
K^{(2)}(z,0)&=&P_{eq}(\beta,z,0)\delta f(\beta,z,0),
\end{eqnarray}
respectively, where the classical canonical distribution is
\begin{eqnarray}
\label{classicaldistribution}
P_{eq}(\beta, z,0)=\frac{e^{-\beta H(z,0)}}{Z_C(0)},
\end{eqnarray}
$Z_C(0)$ is the classical partition function of the system at time 0,
\begin{eqnarray}
\delta f(\beta,z,0)=f(\beta,z,0)-\langle f(0)\rangle_{eq},
\end{eqnarray}
and $\langle  f(0) \rangle_{eq}$ is an average of $f(\beta,z,0)$ with respect to the canonical distribution~(\ref{classicaldistribution}). Here, we explicitly mark the parameter $\beta$ since we replace it by other parameters in the next section. Eq.~(\ref{initialconditionK2}) originates from the quantum corrections to the classical distribution~\cite{Wigner1932}.
We immediately find that Eq.~(\ref{equationmotioninphasespace0order}) (the zeroth power of $\hbar$) is nothing but the celebrated FK formula for the classical work~\cite{Jarzynski1997a,Hummer2001,Imparato2005,Ge2008,Chetrite2008,Liu2009}
\begin{eqnarray}
W=\int_0^t \partial_s H[z(s),s] ds,
\end{eqnarray}
and the solution is
\begin{eqnarray}
\label{K0solution}
K^{(0)}(z,t)&=&\left\langle e^{i\eta\int_0^t \partial_s H[ z(s),s] ds }\delta(z-z(t))\right\rangle \nonumber \\
%&=&\int dz'G(z,t|z',0)P_{eq}(\beta,z',0)\nonumber\\
&=&e^{i\eta[H(z,t)-H(\psi_0^{-1}(z,t),0)]}P_{eq}(\beta, \psi_0^{-1}(z,t),0).
\end{eqnarray}
The angular brackets indicate an average over all classical phase trajectories that started from the initial canonical distribution weighted by the exponential work. The second equation is valid only for closed classical systems, where $\psi_0^{-1}$ is the inverse of the flow map of the classical Hamiltonian system,
\begin{eqnarray}
\label{flowmap}
z=\psi_t(z_0,0).
\end{eqnarray}
That is, the phase points $z_0$ at time 0 and $z$ at time $t$ are on the same phase trajectories connected by the map $\psi_t$. Therefore, a quantum-classical correspondence of the FK formula is explicitly established. Obviously, the corresponding principle of the work statistic is then a natural consequence. This is the first important result obtained in this paper.

Given the flow map~(\ref{flowmap}), we can also easily construct the solution of Eq.~(\ref{equationmotioninphasespace1order}) (the first power of $\hbar$):
\begin{eqnarray}
\label{solutionK1}
K^{(1)}(z,t)%&=&\left\langle\delta(z-z(t))\int_0^t ds e^{i\eta\int_0^s \partial_s H(z(s),s) ds }\frac{i}{2}\left\{ \eta^2 \left[ \partial_{s} H(z(s),s) \Lambda H(z(s),s) \right] + i\eta\partial_{s} H(z(s),s) \Lambda\right\} K_0(z(s),s)\right\rangle \nonumber \\
%&=&\left\langle \int_0^t dt' \frac{i}{2}\left[ \eta^2 \left( {\partial_s H(z(s),s)} \Lambda H(z(s),s) \right) -i\eta {\partial_s H(z(s),s)} \Lambda\right] K_0(z(s),s) \right\rangle\nonumber \\
%&=&-\int_0^t dt' \int dz' G(z,t|z',t')\frac{i}{2}\left\{ \eta^2 \left[ \partial_{t'} H(z',t') \Lambda H(z',t') \right] + i\eta\partial_{t'} H(z',t') \Lambda\right\} K_0(z',t')\nonumber \\
&=&\int_0^t dt' e^{i\eta[H(z,t)-H(z',t')]} \frac{1}{2}\left\{ (i\eta)^2 \left[ {\partial_{t'} H(z',t')} \Lambda' H(z',t' \right)  ]- \right.\nonumber \\
&&\hspace{0cm}\left. i\eta \partial_{t'} H(z',t')\Lambda'\right\} K_0(z',t')\left|_{z'=\psi_{t'}^{-1}(z,t)},\right.
\end{eqnarray}
where $z'$ is the phase point at time $t'$, which is connected to the phase point $z$ at time $t$, namely, $z=\psi_t(z',t')$. The reader is reminded that the operator $\Lambda'$ in this equation is defined with respect to $z'=(x',p')$. A very analogous expression for $K^{(2)}(z,t)$ can also be obtained, and it obviously depends on the initial condition~(\ref{initialconditionK2}) in addition to functions $K^{(0)}$ and $K^{(1)}$.
%\begin{eqnarray}
%K_2(z,t)&=&\int dz'G(z,t|z',0)K_2(z',0)-\int_0^t dt' dz' G(z,t|z',t')\frac{i}{2}\left\{ \eta^2 \left[ \partial_{t'} H(z',t') \Lambda H(z',t') \right] %+ \right. \nonumber \\
%&&\left. i\eta\partial_{t'} H(z',t') \Lambda\right\} K_1(z',t') + \int_0^t dt' dz' G(z,t|z',t')\left(\frac{i}{2}\right)^2(\cdots K_0(z')\cdots)  %\nonumber\\
%&=&e^{i\eta[H(z,t)-H(\psi_0^{-1}(z,t),0)]}P_{eq}(\psi_0^{-1}(z,t),0),
%\end{eqnarray}
Based on these above observations, we arrive at the series expansion of the CF~(\ref{CFworkclosedsystem}) in powers of $\hbar$ as follows:
\begin{eqnarray}
\Phi(\eta)=\Phi^{(0)}(\eta) + (i\hbar)\Phi^{(1)}(\eta)+(i\hbar)^2\Phi^{(2)}(\eta) +\cdots,
\end{eqnarray}
where
\begin{eqnarray}
\label{CF0thorder}
\Phi^{(0)}(\eta)%&=&\int dz K^{(0)}(z,t) %e^{i\eta [H(\psi_t(z,0),t)-H(z,0)]}P_{eq}(z,0)%=\left\langle e^{i\eta W}\right\rangle,\\
&=&\left\langle e^{i\eta\int_0^t \partial_s H(z(s),s) ds }\right\rangle,  \\
\label{CF1storder}
\Phi^{(1)}(\eta)%&=&\int dz K^{(1)}(z,t)\nonumber\\%-\int_0^t dt'\int dz' e^{i\eta[H(\psi_t(z',t'),t)-H(z',t')]} \frac{i}{2}\left\{ \eta^2 \left[ {\partial_{t'} H(z',t')} \Lambda H(z',t'\right) ]+ \right.\nonumber \\
%&&\left. i\eta \partial_{t'} H(z',t')\Lambda\right\}K_0(z',t').\nonumber\\
&=&\frac{(i\eta)(i\eta+\beta)}{2}\int dz e^{i\eta [H(\psi_t(z,0),t)-H(z,0)]}P_{eq}(z,0)\int_0^t dt'\partial_{t'}H(\psi_{t'}(z,0),t')\Lambda H(z,0)\nonumber\\
&=&\frac{(i\eta)(i\eta+\beta)}{2}\left\langle  e^{i\eta\int_0^t \partial_s H(z(s),s) ds }\int_0^t ds\partial_{s}H(z(s),s)\Lambda H(z(0),0)\right\rangle.
%&=&\int dz e^{i\eta[H(\psi_t(z,0),t)-H(z ,0)]} \int_0^t dt'\frac{i}{2}\left\{ \eta^2 \left[ {\partial_{t'} H(\psi_{t'}(z,0),t')} \Lambda H(\psi_{t'}(z,0),t'\right)  ]-\right.\nonumber \\
%&&\left. i\eta \partial_{t'} H(\psi_{t'}(z,0),t),t')\Lambda\right\}P_{eq}(z,0).
\end{eqnarray}
In the derivations of these equations, we used the Liouville theorem. In addition, we did not write $\Phi^{(2)}(\eta)$ temporally since its current form is too long to be useful.

\section{Lowest order quantum correction}
\label{section4}
As we noted at the beginning, the ultimate goal of studying the quantum-classical correspondence of work statistics is to deepen our understanding of the quantum characteristics of work. Hence, we are interested in the quantum corrections of the CF with higher orders of $\hbar$, e.g., $\Phi^{(1)}(\eta)$ and above. However, Eq.~(\ref{CF1storder}) implies that the work moments, which are calculated by taking different orders of derivatives of the CF with respect to $i\eta$, are complex numbers. Hence, $\Phi^{(1)}(\eta)$ must be zero. This fact is not very apparent if we simply look at Eq.~(\ref{solutionK1}). After carefully revisiting Eqs.~(\ref{equationmotioninphasespace0order})-(\ref{equationmotioninphasespace1order}), we find that there is a key relation between their solutions:
\begin{eqnarray}
\label{keyidentity}
K^{(1)}=-\frac{i\eta}{2} H\Lambda K^{(0)}.
\end{eqnarray}
Obviously, the above equation ensures that the first order quantum correction $\Phi^{(1)}$ is exactly zero. As a result, if one wants to obtain meaningful quantum corrections, we must expand the quantum CF at least to the second order of $\hbar$. In principle,  Eq.~(\ref{equationmotioninphasespace2order}) has provided the answer. However, this equation is too complicated to solve.

In fact, the quantum CF~(\ref{CFworkclosedsystem}) has an alternative expression,
\begin{eqnarray}
\label{CFworkclosedsystem2}
\Phi(\eta)&=&{\rm Tr}\left[ e^{i\eta\hat{H}(t)} U(t)e^{-i\eta \hat{H}(0)}\rho_0U^\dag(t)\right]\equiv {\rm Tr}\left[ e^{i\eta\hat{H}(t)} \hat{\rho}(t)\right].
\end{eqnarray}
We call $\hat \rho(t)$ the heat characteristic operator (HCO)~\cite{Liu2018}. The operator satisfies the Liouville-von Neumann equation
\begin{eqnarray}
\label{vonNeumannequation}
{\partial_t}\hat{\rho}(t)  =&&\frac{1}{i\hbar}[\hat{H}(t),\hat{\rho}(t)],
 \end{eqnarray}
with the modified initial condition
\begin{eqnarray}
\hat\rho(0)=\frac{e^{-(i\eta+\beta) \hat H(0)}}{Z(0)}.
\end{eqnarray}
We may write Eq.~(\ref{vonNeumannequation}) in the phase space representation as well: let the Weyl symbol $[\hat\rho(t)]_w=P(z,t)$; then, the following equation is satisfied,
\begin{eqnarray}
\label{equationmotioninphasespacetillsecondorder2}
 {\partial_t} P(z,t)=-H(z,t)\Lambda P(z,t) +   (i\hbar)^2 \frac{1}{24}{\partial_x^3 }U {\partial_p^3 }P(z,t)+\cdots, %\sum_{n=2} {\cal O}(\hbar^{2n}),
\end{eqnarray}
and the initial condition is
\begin{eqnarray}
\label{initialconditionHCOWeylsymbol}
P(z,0)=P_{eq}(i\eta+\beta,z,0)+ (i\hbar)^2P_{eq}(i\eta+\beta,z,0)\delta f( i\eta+\beta,z,0)+\cdots.
\end{eqnarray}
The reader is reminded that Eqs.~(\ref{equationmotioninphasespacetillsecondorder2}) and~(\ref{initialconditionHCOWeylsymbol}) contain only terms of even powers of $\hbar$. If one can solve Eq.~(\ref{equationmotioninphasespacetillsecondorder2}), the CF is evaluated by
\begin{eqnarray}
\label{evaluatingPhi2}
\Phi(\eta)=\int_{-\infty}^{+\infty} dz \{\exp[i\eta \hat H(t)]\}_wP(z,t).
\end{eqnarray}
In general, to solve Eq.~(\ref{equationmotioninphasespacetillsecondorder2}) is a very difficult task. Hence, we have to resort to the $\hbar$ series expansion again.
%%%Quality Control Editor - Please ensure that the intended meaning has been maintained in the above edit.
Expanding the solution of Eq.~(\ref{equationmotioninphasespacetillsecondorder2}) in even powers of $\hbar$~\footnote{The reason for this is that the evolution equation and initial condition contain only even powers of $\hbar$},
\begin{eqnarray}
\label{seriesexpansionofPfunction}
P(z,t)=P^{(0)}(z,t) + (i\hbar)^2 P^{(2)}(z,t) +\cdots% +\sum_{n=2} {\cal O}(\hbar^{2n}),
\end{eqnarray}
we obtain the following solutions:
\begin{eqnarray}
P^{(0)}(z,t)=&&P_{eq}[i\eta+\beta,\psi^{-1}_0(z,t),0], \\
P^{(2)}(z,t)=&&P_{eq}[i\eta+\beta,\psi^{-1}_0(z,t),0]\hspace{0.1cm}\delta f[i\eta+\beta,\psi^{-1}_0(z,t),0] +\nonumber\\
&& \frac{1}{24}\int_0^t dt'{\partial_{x'}^3 }U \partial_{p'}^3 P^{(0)}(z',t') \left|_{z'=\psi^{-1}_{t'}(z,t)}\right..
\end{eqnarray}
According to Eq.~(\ref{evaluatingPhi2}), $P^{(0)}(z,t)$ obviously gives the zeroth-order CF, Eq.~(\ref{CF0thorder}). If we substitute $P^{(2)}(z,t)$ and collect all terms of the second power of $\hbar$, we find that the quantum correction of the second order of $\hbar$ is composed of three terms,
\begin{eqnarray}
\label{CF2ndorder}
\Phi^{(2)}(\eta)=\Phi^{(2)}_m(\eta)+ \Phi^{(2)}_i(\eta) + \Phi^{(2)}_d(\eta),
\end{eqnarray}
%\begin{eqnarray}
%\Phi_C(\eta)&=&\int dz e^{i\eta H(z,t)}P^{(0)}(z,t)%=\int dz_0  e^{i\eta[ H(\phi_t(z_0,0),t)-H(z_0,0)]}P_{eq}(z_0,0) \nonumber\\
%=\left\langle  e^{i\eta\int_0^t ds \partial_s H( z(s),s) } \right\rangle
%\end{eqnarray}
where
\begin{eqnarray}
\label{measureeffect}
\Phi^{(2)}_m(\eta)%\int dz e^{i\eta H(z,t)} f(-i\eta,z,t)  P^{(0)}(z,t)\nonumber]\\
&=&\left\langle e^{i\eta\int_0^t ds \partial_s H[z(s),s] }   f[-i\eta,z(t),t] \right\rangle, \\
\label{initialeffect}
\Phi^{(2)}_i(\eta)%&=&\int dz e^{i\eta H(z,t)}P_{eq}(i\eta+\beta,\psi^{-1}_0(z,t),0)\delta f[ i\eta+\beta,\psi^{-1}_0(z,t),0] \nonumber\\
&=&\left\langle   e^{i\eta\int_0^t ds \partial_s H[z(s),s] }  \delta f[i\eta+\beta,z(0),0] \right\rangle, \\
\label{dynamiceffect}
\Phi^{(2)}_d(\eta)%&=&\frac{1}{24}\int dz e^{i\eta H(z,t)}P_{eq}(i\eta+\beta,\psi^{-1}_0(z,t),0)\int_0^t dt' {\partial_{x'}^3 }U  {\partial_{p'}^3 }P^{(0)}(z',t') \left|_{z'=\psi^{-1}_{t'}(z,t)}\right. \nonumber\\
&=&\left\langle   e^{i\eta\int_0^t ds \partial_s H[z(s),s] } \int_0^t ds Q [ z(s),s,i\eta+\beta] \right\rangle.
\end{eqnarray}
The integrand in the last equation is
\begin{eqnarray}
%Q(z,s,i\eta+\beta)=
\frac{1}{24}{\partial_x^3}U\left[ -(i\eta+\beta){\partial_p^3}\widetilde H+3(i\eta+\beta)^2 ({\partial_p^2}\widetilde H)( {\partial_p}\widetilde H)-(i\eta+\beta)^3 \left( \partial_p \widetilde H \right)^3\right],
\end{eqnarray}
and
\begin{eqnarray}
\widetilde H(z,s)\equiv H[\psi_0^{-1}(z,s),0].
\end{eqnarray}
Although these terms seem complicated in form, particularly Eq.~(\ref{dynamiceffect}), their physical origins are very clear: $\Phi^{(2)}_m(\eta)$ is the quantum effect of the second energy projective measurement, $\Phi^{(2)}_i(\eta)$ arises from the quantum correction of the initial condition, and $\Phi^{(2)}_d(\eta)$ is the quantum correction to the classical dynamical equation. Therefore, these quantum effects manifest themselves independently in the corrections of the second power of $\hbar$. Before closing this section, we want to present two comments. One is that Eq.~(\ref{keyidentity}) has a simple explanation based on the dynamics about $P(z,t)$; see Appendix B. The other is the observation that quantum corrections to the classical CF of work only include terms with even powers of $\hbar$. The reason is obvious if one notes that all $P^{(n)}(z,t)$ with odd powers of $\hbar$ are exactly zero; see Eq.~(\ref{seriesexpansionofPfunction}). This is the second important result in this paper.

\section{Two examples}
\label{section5}
\subsection{A forced harmonic oscillator}
We use a driven harmonic oscillator to analytically illustrate these quantum corrections, where the Hamiltonian of the system is simply
\begin{eqnarray}
\hat{H}=\frac{\hat p^2}{2m}+\frac{m\omega^2\hat x^2}{2} +  F(t)\hat x,
\end{eqnarray}
where $\omega$ is the angular frequency, and $F(t)$ is the external driving force, which we assume to be zero at time $0$. The quantum CF of this system has an analytical formula~\cite{Talkner2008}:
\begin{eqnarray}
\label{CFharmonicoscillator}
\Phi_{HO}(\eta)=\exp\left[-\frac{i\eta F(t)^2 }{2m\omega^2}+ c(t)\frac{\left(e^{i\eta\hbar\omega} -1\right)}{\hbar\omega} -4 c(t) \frac{\sin(\hbar\omega\eta/2)^2}{\hbar\omega\left(e^{\beta\hbar\omega}-1\right)}   \right],
\end{eqnarray}
where
\begin{eqnarray}
c(t)=\frac{1}{2m\omega^2}\left| \int_0^t ds \dot{F} e^{i\omega s}\right |^2,
\end{eqnarray}
and the dot denotes a derivative with respect to time. We use the subscript `$HO$' to denote that it is the exact of the quantum harmonic oscillator. Expanding Eq.~(\ref{CFharmonicoscillator}) in powers of $\hbar$ to the second order, we have
\begin{eqnarray}
\label{expand0thorderexact}
\Phi_{HO}^{(0)}(\eta)&=&\exp\left[-\frac{i\eta F(t)^2 }{2m\omega^2}+ i\eta c(t) - \frac{\eta^2 c(t)}{\beta}\right],\\
\label{expand2ndorderexact}
\Phi_{HO}^{(2)}(\eta)&=&\frac{(\beta+i\eta)^2\eta^2\omega^2c(t)}{12\beta}\Phi_{HO}^{(0)}(\eta).
\end{eqnarray}
Moreover, if we expand the quantum CF (Eq.~(\ref{CFharmonicoscillator})) in higher powers of $\hbar$, we can easily verify that there are only $\hbar^{2n}$-terms. Hence, the model of the forced harmonic oscillator confirms our results.

Now we are in position to check whether Eqs.~(\ref{CF0thorder}) and~(\ref{CF2ndorder}) can be used to reproduce Eqs.~(\ref{expand0thorderexact}) and~(\ref{expand2ndorderexact}), respectively. Using Eq.~(\ref{K0solution}) and the exact flow map of the classical harmonic oscillator (see Appendix C), we can straightforwardly calculate the zeroth-order CF, $\Phi^{(0)}(\eta)$, and the result agrees with Eq.~(\ref{expand0thorderexact}). The calculation of $\Phi^{(2)}(\eta)$ is relatively complicated. Because the potential of the harmonic oscillator is
\begin{eqnarray}
U(x,t)=\frac{m\omega^2x^2}{2}+F(t)x,
\end{eqnarray}
the dynamical correction term, $\Phi_d^{(2)}(\eta)$, is zero. In addition, in this specific system,
\begin{eqnarray}
\langle f(0)\rangle_{eq}=\frac{\omega^2\beta^2}{24}.
\end{eqnarray}
Substituting all relevant quantities into Eqs.~(\ref{measureeffect}) and~(\ref{initialeffect}) and after some algebraic calculations, we obtain the two corrections as follows:
\begin{eqnarray}
\Phi^{(2)}_i(\eta)&=&\left[- \frac{\omega^2(i\eta+\beta)^3(2\beta-2\eta^2 c(t))}{24\beta^2}+ \frac{\omega^2(i\eta+\beta)^2}{8}-\frac{\omega^2\beta^2}{24}\right ] \Phi_{HO}^{(0)}(\eta).\\
\Phi^{(2)}_m(\eta)&=&\left[\frac{\omega^2(i\eta)^3(2\beta-2\eta^2 c(t))}{24\beta^2}+\frac{(i\eta)^4\omega^2 c(t)}{6\beta}+ \frac{\omega^2(i\eta)^3c(t)}{12}-\frac{\omega^2\eta^2}{8}\right] \Phi_{HO}^{(0)}(\eta).
\end{eqnarray}
Their sum is exactly $\Phi^{(2)}_{HO}(\eta)$ given in Eq.~(\ref{expand2ndorderexact}). Some useful formulas in the derivation are presented in Appendix~C.

\begin{figure}
\includegraphics[width=1.\columnwidth]{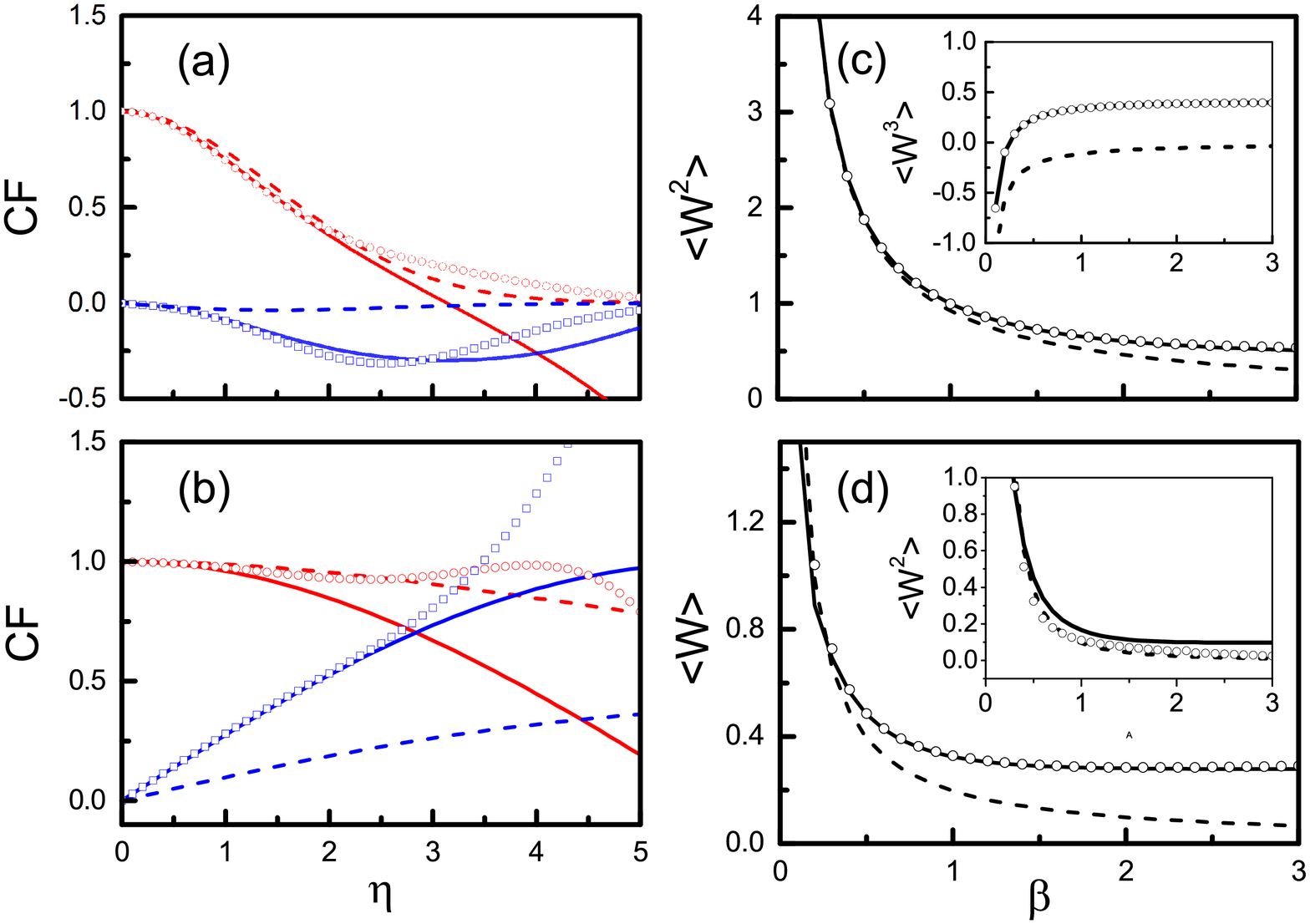}
\caption{Real (in red) and imaginary (in blue) parts of the CFs of the forced harmonic oscillator (a) and the forced quartic oscillator (b) at $\beta=2$. The solid and dashed lines are the results of the quantum and classical CFs, respectively. The circles and squares are the results of the clasical CFs with the $\hbar^2$-quantum corrections. (c) and (d) and their insets are the work moments versus the inverse temperature $\beta$ for the forced harmonic oscillator and the forced quartic oscillator, respectively. They are evaluated by using these CFs. We do not show the first work moment (or the mean work) for the harmonic oscillator since these CFs give the same results for the special model. We set $m=\hbar=\omega=1$ for the harmonic oscillator, while $m=1/2$ for the quartic oscillator. }
\label{figure1}
\end{figure}

To explicitly show the importance of the $\hbar^2$-quantum correction, we show the CFs in panel (a) of Fig.~\ref{figure1}, which includes the exact quantum CF, the classical CF, and the classical CF with the quantum corrections. We apply a linear force, $F(t)=t$ ($0\le t \le 1$) therein. Although the semiclassical CF cannot completely recover the exact one, particularly at large $\eta$ values in lower temperature case, it indeed improves in comparison with the classical CF. If we check their work moments, this improvement is more prominent; see panels (c) in the same figure. We also would like to emphasize our scheme is not restricted to the specific example. To illustrate this, we calculate the CFs for a forced quartic oscillator as a second example.

\subsection{A forced quartic oscillator}
Here we want to consider a forced quartic oscillator~\cite{Jarzynski2015}, whose the Hamiltonian is
\begin{eqnarray}
\hat{H}=\frac{\hat p^2}{2m}+F(t)\hat x^4,
\end{eqnarray}
where we apply $F(t)=1+t$. The model is distinct from the harmonic oscillator since in this case the dynamical correction~(\ref{dynamiceffect}) does not vanish. And to our knowledge, the analytical expression of the quantum CF is not available. Hence, we have to do numerical simulations. Data of the quartic oscillator are shown in Fig.~\ref{figure1}, (b) and (d) show the data. We see again that the quantum correction terms are significant in bridging the quantum and the classical CFs in lower temperature case.

\section{Conclusion}
\label{section6}
In this paper, we studied the quantum corrections of the work statistics in closed quantum systems by expanding the quantum CF of work in powers of $\hbar$. The forced harmonic oscillator and the forced quartic oscillator clearly verity the validity of the our formulas, particularly in the range of moderate and high temperatures. We think that our results will be useful when studying complicated quantum systems. The phase trajectory of classical system and the thermal equilibrium state can be efficiently simulated by molecular dynamics and/or Monte-Carlo methods, so $\hbar^2$ corrections provide a rigorous alternative to full quantum work statistics.

%Although in principle the semiclassical approximation can be realized by the other techniques, e.g., WKB approximation or the Feynman path integral method. This paper provides an new routine to the semi-classical description of the quantum work statistics. These methods are expected to give equivalent results.
There are several possible theoretical extensions of the current work. For instance, if we take into account higher powers of $\hbar$, it should be interesting to see whether these additional quantum corrections can lead to significant improvements of the quantum CF of work. In addition, if there are many identified particles in quantum systems, quantum statistics have to be taken into account. Finally, for open quantum systems, we have established the quantum FK formula as well~\cite{Liu2014,Liu2014a,Liu2018}. The exact meaning of quantum-classical correspondence in these situations is worth investigating in detail.\\
\\

\begin{acknowledgments}
This work was supported by the National Science Foundation of China under Grant Nos. 11174025 and 11575016. We also appreciate the support of the CAS Interdisciplinary Innovation Team, No. 2060299.
\end{acknowledgments}

\section*{Appendix A: the second power of $\hbar$ in Eq.~(\ref{equationmotioninphasespacetillsecondorder1})}
The term $(\cdots K\cdots)$ is complicated since it includes the contributions from both the commutator $[,$ $]$ and $\hat\Omega$:
\begin{eqnarray}
%\left(\frac{i\hbar}{2 }\right)^2
(\cdots K\cdots)&=&\frac{1}{24}\frac{\partial^3 U}{\partial x^3 } \frac{\partial^3 K}{\partial p^3 } +\left\{\frac{i\eta}{8}\left( \frac{\partial H}{\partial t}e^{i\eta H} \right)\Lambda^2\left( e^{-i\eta H}\right) \right.+\nonumber\\
&&\left. i \eta \frac{\partial  H}{\partial t} \left[ f(z,t,i\eta) + f(z,t,-i\eta)\right]+\frac{\partial}{\partial t}f(z,t,-i\eta)
- \frac{(i\eta)^2}{4} \left(\frac{\partial H}{\partial t } \Lambda H\right)\Lambda + \frac{i \eta}{8}\frac{\partial H}{\partial t} \Lambda^2 \right\}K,
\end{eqnarray}
where
\begin{eqnarray}
\Lambda^2=\overleftarrow{{\partial_p^2}}\overrightarrow{ {\partial_x^2}} - 2\overleftarrow{{\partial_x\partial_p}}\overrightarrow{ {\partial_p\partial_x}} -\overleftarrow{ {\partial_x^2}}\overrightarrow{ {\partial_p^2}}.
\end{eqnarray}

\section*{Appendix B: an alternative understanding of Eq.~(\ref{keyidentity}) }
According to the definitions of $\hat K$ and $\hat \rho$, we have
\begin{eqnarray}
\hat \rho= e^{i\eta\hat H(t)} \hat K.
\end{eqnarray}
Expressing them in the phase space representation and expanding them to the second power of $\hbar$, we have
\begin{eqnarray}
&&P^{(0)}(z,t) + (i\hbar)^2 P^{(2)}(z,t)+\cdots\nonumber \\
=&&e^{-i\eta H(z,t)}K^{(0)}(z,t) + (i\hbar) e^{-i\eta H(z,t)}\left[ K^{(1)}+ \frac{i\eta}{2}H(z,t)\Lambda K^{(0)}(z,t)\right] + \dots
\end{eqnarray}
The term proportional to $\hbar$ on the left hand side is zero, so we immediately reobtain Eq.~(\ref{keyidentity}). Of course, this result is imposed by the Liouville-von Neumann equation and the specific initial condition.
%%%Quality Control Editor - Please ensure that the intended meaning has been maintained in the above edit.

\section*{Appendix C: several useful formulas for the forced harmonic oscillator}
The flow map $\psi$ of the classical harmonic oscillator with the Hamiltonian
\begin{eqnarray}
H(z,t)=\frac{p^2}{2m}+\frac{m\omega^2x^2}{2} + F(t)x
\end{eqnarray}
has the following analytical expressions:
\begin{eqnarray}
x(t)&=&x_0 \cos(\omega t) +\frac{p_0}{m\omega}\sin(\omega t) - l(t),\\
p(t)&=&-m \omega x_0 \sin(\omega t)+p_0 \cos(\omega t) - \dot{l}(t),
\end{eqnarray}
where the function $l(t)$ is
\begin{eqnarray}
l(t)=\frac{1}{m\omega}\int_0^t F(s)\sin(\omega(t-s))ds.
\end{eqnarray}
Hence, the difference of the Hamiltonian at two phase points along the same phase trajectory is
\begin{eqnarray}
H(z(t),t)-H(z_0,0)=a(t)x_0+b(t)p_0+c(t)-\frac{F(t)^2}{2m\omega^2},
\end{eqnarray}
where
\begin{eqnarray}
a(t)&=&m\omega \sin(\omega t)\dot{l}(t)-m\omega^2\cos(\omega t)l(t)+F(t)\cos(\omega t),\\
b(t)&=&-\cos(\omega t)\dot{l}(t) -\omega \sin(\omega t)l(t)+\frac{F(t)}{m\omega}\sin(\omega t).
\end{eqnarray}
To determine these results, we used the following relation:
\begin{eqnarray}
c(t)&=&\frac{[F(t)-m\omega^2l(t)]^2}{2m\omega^2} +\frac{m\dot{l}(t)^2}{2}=\frac{mb(t)^2}{2}+\frac{a(t)^2}{2m\omega^2}.
\end{eqnarray}
These equations are used in deriving the concrete expressions of $\Phi^{(2)}_i(\eta)$ and $\Phi^{(2)}_m(\eta)$.

\bibliography{RFsubmission2018330}.
\iffalse
%
\fi

\end{document}